\pdfoutput=1

\documentclass[10pt,conference]{IEEEtran}
\IEEEoverridecommandlockouts
\usepackage{cite}
\usepackage[left=0.625in,right=0.625in,top=0.65in,bottom=1in]{geometry}
\usepackage[T1]{fontenc}
\usepackage{cite}
\usepackage{graphicx}
\usepackage{array}
\newcolumntype{L}{>{\raggedright\arraybackslash}p{3cm}}
\usepackage{epstopdf}
\usepackage{comment}
\usepackage{url}
\usepackage{xcolor}
\usepackage{algorithmicx}
\usepackage[colorlinks=true,allcolors=blue,bookmarks=false]{hyperref}
\ifCLASSINFOpdf
\else
\fi
\usepackage{amsmath}

\usepackage{color}

\usepackage{amsfonts}
\usepackage{subfigure}
\usepackage{amsmath}

\usepackage[ruled,vlined]{algorithm2e}
\definecolor{seagreen}{rgb}{0.18, 0.55, 0.24}
\SetAlFnt{\small\color{black}\tt}

\DeclareMathAlphabet{\mathbit}{OT1}{cmr}{bx}{it}
\usepackage[cmintegrals]{newtxmath}

\begin{document}

\title{ Lightweight AI for UAV-Mounted RIS: An Overview\thanks{This work was supported by JSPS KAKENHI Grant Numbers JP25K07743 and JP23K24905, Japan}}

 \author{
\IEEEauthorblockN{Sherief Hashima\IEEEauthorrefmark{1}\IEEEauthorrefmark{2}, Kohei Hatano \IEEEauthorrefmark{1}\IEEEauthorrefmark{3}, Eiji Takimoto \IEEEauthorrefmark{3},
Mohamed Rihan\IEEEauthorrefmark{4}\IEEEauthorrefmark{5}, Basem.M. Elhalawany \IEEEauthorrefmark{6} \IEEEauthorrefmark{7},\\ and 
 Hamada Rizk\IEEEauthorrefmark{8} 
}

\small
\IEEEauthorblockA{\IEEEauthorrefmark{1}Computational Learning Theory Team, RIKEN-AIP, Fukuoka, 819-0395, Japan.\\
\IEEEauthorrefmark{2} Engineering Dept, Nuclear Research center, Egyptian Atomic Energy Authority, 13759, Cairo, Egypt.}
\IEEEauthorblockA{\IEEEauthorrefmark{3}Department of Informatics, Kyushu University, Fukuoka 819-0395, Japan.}
\IEEEauthorblockA{\IEEEauthorrefmark{4} Department of Communications Engineering, University of Bremen, Bremen 28359, Germany}
\IEEEauthorblockA{\IEEEauthorrefmark{5} \small{Electronics and Electrical Commun. Engineering, Faculty of Electronic Engineering, Menoufia University, Menouf 32952, Egypt}}
\IEEEauthorblockA{\IEEEauthorrefmark{6}
Kuwait College of Science and Technology,
Aljahraa, Kuwait} 
 \IEEEauthorblockA{\IEEEauthorrefmark{7} Department of Electrical Engineering, Faculty of Engineering at Shoubra, Benha University, Cairo, Egypt.}
 \IEEEauthorblockA{\IEEEauthorrefmark{8} The University of Osaka, Osaka, Japan}
}
\maketitle
\begin{abstract}
Unmanned Aerial Vehicles (UAV)-mounted Reconfigurable Intelligent Surfaces (RIS) have emerged as a promising architecture for enhancing wireless coverage, spectral efficiency, and energy performance in 6G networks. By combining programmable electromagnetic wave manipulation with aerial mobility, UAV-RIS systems enable dynamic blockage mitigation, adaptive beamforming, and flexible deployment across terrestrial, maritime, and satellite-integrated environments. However, joint optimization of UAV trajectory, RIS phase configuration, and resource allocation incurs high computational complexity, which is incompatible with the strict energy and onboard processing constraints of UAV platforms. Lightweight AI techniques offer practical solutions to this challenge. Hence, this paper provides a comprehensive overview of lightweight AI techniques for UAV-mounted RIS systems, including Reinforcement Learning (RL), meta-learning, Federated Learning (FL), Multi-Armed Bandits (MAB), and energy-aware optimization. We present a detailed taxonomy and comparative analysis of existing work, highlight computational–energy trade-offs, and identify open research challenges for scalable, energy-efficient airborne intelligent surfaces. Furthermore, we present a case study demonstrating the effect of MAB schemes on throughput and energy efficiency in UAV-mounted RIS.

\end{abstract}

\begin{IEEEkeywords}
UAV mounted RIS;  Lightweight AI; RL, MAB, Meta learning;
\end{IEEEkeywords}
\section{Introduction}

Reconfigurable Intelligent Surfaces (RIS) have been recognized as a key enabler of programmable wireless environments for beyond-5G and 6G systems \cite{ref2,ref4}. RIS enables controllable reflection of electromagnetic waves, improving coverage and Energy Efficiency (EE) without requiring active Radio Frequency (RF) chains. While early research focused on static, ground-mounted RIS, recent developments have introduced UAV-mounted RIS systems, where programmable reflecting surfaces are integrated into aerial platforms \cite{ref5,ref6,ref7,ref8,ref9}.

UAV-mounted RIS systems provide unique advantages over terrestrial RIS installations. By leveraging 3D mobility, UAV-RIS can dynamically adjust both horizontal position and altitude to optimize channel conditions \cite{ref10,ref11,ref12,ref13}. This flexibility enables rapid deployment in emergency scenarios \cite{ref6}, extended connectivity in rural or maritime environments \cite{ref12}, and integration with satellite and Non-Terrestrial Networks (NTN) \cite{ref17,ref18,ref19,ref44,ref44b}. Also, UAV-RIS has been explored for Integrated Sensing and Communication (ISAC) \cite{ref20}, secure communication \cite{ref30}, and Mobile Edge Computing (MEC) \cite{ref26,ref27}.

\begin{figure}[!htbp]
\centering
\includegraphics[height=0.5\columnwidth]{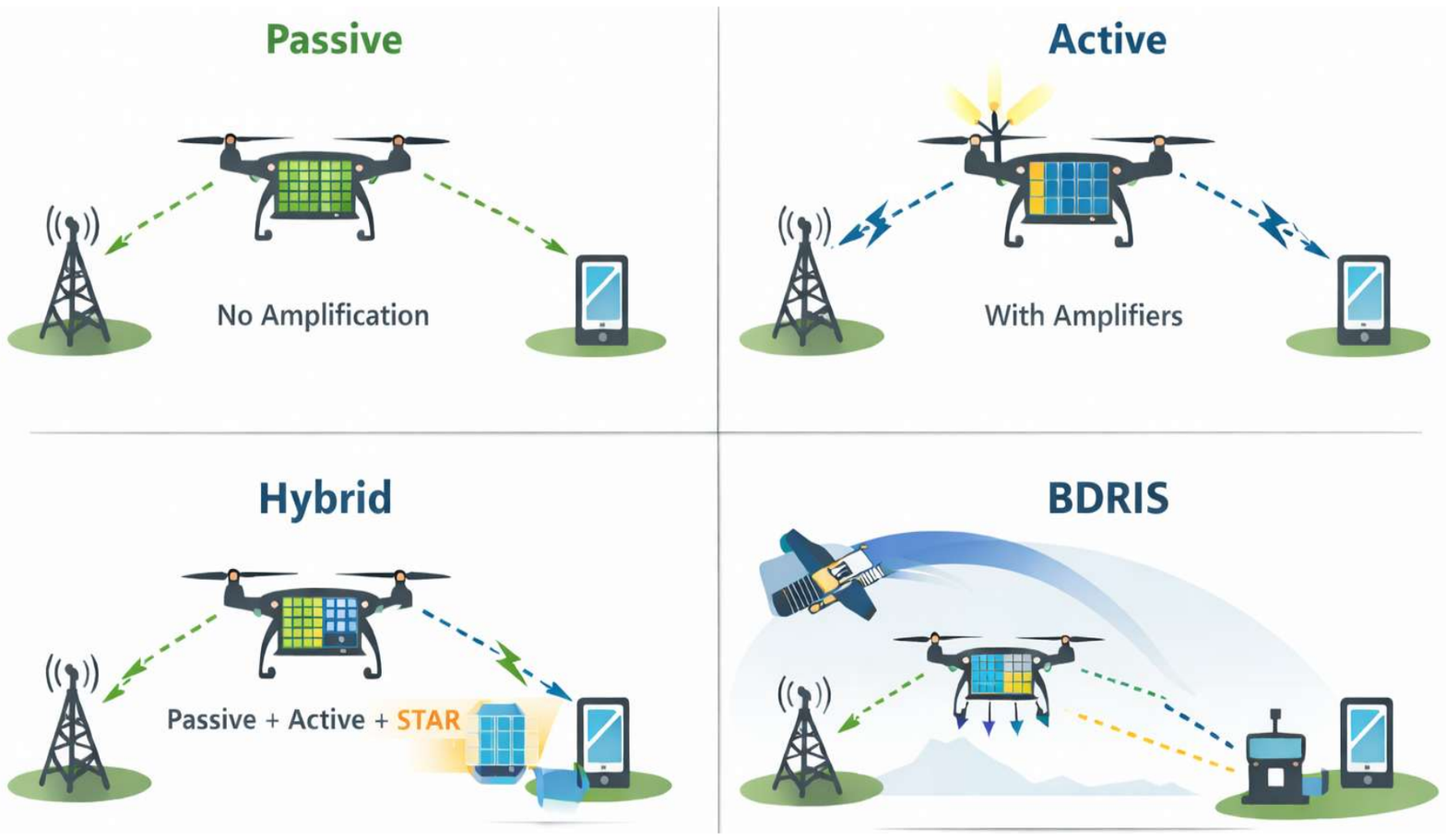}
\vspace{-5mm}
\caption{UAV-mounted RIS architectures}
\label{ristypes}
\end{figure}

Moreover, recent advancements in RIS hardware architectures have introduced multiple RIS types with distinct operational characteristics, as highlighted in Table \ref{tab:RISarchs}.  Figure.\ref{ristypes} presents the four main types of aerial RIS used in different scenarios. Conventional passive RIS relies solely on controllable Phase Shifts (PSs) without signal amplification, offering low power consumption and hardware simplicity \cite{ref4}. Furthermore, active RISs incorporate low-power amplifiers within reflecting elements to enhance signal strength, albeit at the cost of increased energy consumption and hardware complexity \cite{ref13,ref30}. Simultaneously Transmitting and Reflecting RIS (STAR-RIS) architectures enable simultaneous signal transmission and reflection to serve users on both sides of the RIS panel \cite{ref8,ref17}. This significantly improves spatial coverage and flexibility in aerial RIS-assisted deployments. Furthermore, hybrid RIS architectures that combine passive, active, and STAR components aim to balance EE and performance enhancement \cite{ref15,ref33}. Emerging Beyond-Diagonal RIS (BD-RIS) and coupled-element designs introduce non-diagonal scattering matrices that enable more flexible wave manipulation and spatial multiplexing, further expanding the design space of airborne RIS systems \cite{ref44,ref44b}. These diverse RIS architectures directly affect the computational and energy requirements of UAV-mounted RIS optimization, underscoring the importance of lightweight AI techniques for real-time control.  
\begin{table*}[!htbp]
\caption{Comparison of RIS Architectures and Their Lightweight AI Requirements for UAV-Mounted Systems}
\label{tab:RISarchs}
\centering
\resizebox{\textwidth}{!}{
\begin{tabular}{|L|L|L|L|L|L|}
\hline
\textbf{RIS Type} 
& \textbf{Operation Mode} 
& \textbf{Power \& Hardware Cost} 
& \textbf{Beamforming Capability} 
& \textbf{Optimization Complexity} 
& \textbf{Typical AI Approach} \\ \hline

\textbf{Passive RIS} \cite{ref4} 
& Reflection only 
& Low 
& Moderate (double fading effect) 
& Moderate 
& RL / MAB \\ \hline

\textbf{Active RIS} \cite{ref13,ref30} 
& Amplified reflection 
& High 
& High gain (attenuation mitigation) 
& High (power + phase control) 
& Energy-aware DRL \\ \hline

\textbf{STAR-RIS} \cite{ref8,ref17} 
& Simultaneous transmit \& reflect 
& Medium–High 
& Dual-side coverage flexibility 
& High (dual-mode control) 
& Meta-RL / DRL \\ \hline

\textbf{Hybrid RIS} \cite{ref15,ref33} 
& Passive + active elements 
& Medium 
& Gain-performance trade-off 
& High (mode selection + amplification) 
& Hybrid RL + model compression \\ \hline

\textbf{BD-RIS} \cite{ref44,ref44b}
& Beyond-diagonal coupled scattering 
& High 
& Enhanced spatial multiplexing 
& Very High (matrix optimization) 
& Deep unfolding / advanced RL \\ \hline

\end{tabular}
}
\end{table*}

Nevertheless, UAV-mounted RIS systems introduce several technical challenges. Mobility and altitude variation lead to time-varying cascaded BS-RIS-user channels and severe Doppler effects \cite{ref9,ref11}. Accurate and frequent Channel State Information (CSI) acquisition becomes costly, particularly when the RIS dimension is large. Moreover, the joint optimization of UAV trajectory, altitude, RIS phase configuration, transmission scheduling, and, in some cases, power allocation or amplification (for active/STAR-RIS) results in a tightly coupled, non-convex, high-dimensional optimization problem \cite{ref22,ref24}. The complexity further increases when advanced RIS architectures such as hybrid or STAR-RIS are employed, as additional transmission/reflection modes must be optimized simultaneously.
From a practical perspective, UAV platforms are inherently constrained by limited battery capacity, strict payload weight limits, and restricted onboard processing capability \cite{ref6,ref21}.  UAV-mounted RIS controllers cannot support iterative convex solvers or large-scale Deep Neural Networks (DNNs) that require extensive floating-point operations. Conventional optimization techniques, including Alternate Optimization (AO) and Successive Convex Approximation (SCA), often involve iterative matrix operations whose complexity scales with the number of RIS elements, making them unsuitable for real-time aerial deployment. As RIS size grows to support higher spatial resolution, computational overhead increases dramatically, exacerbating energy consumption and latency. These limitations motivate the adoption of lightweight AI techniques, rather than generic Deep Learning (DL) frameworks as a necessity for enabling real-time inference under stringent energy and hardware constraints. Techniques such as Reinforcement Learning (RL) and deep RL (DRL) enable model-free decision-making without explicit channel modeling \cite{ref22,ref24}, while meta-learning facilitates rapid adaptation to dynamic channel conditions at reduced retraining cost \cite{ref14}. Federated learning (FL) allows distributed optimization across multiple UAV-RIS nodes without excessive communication overhead \cite{ref31}, and Multi-Armed Bandit (MAB) frameworks provide low-complexity online decision-making suitable for mmWave and dynamic spectrum scenarios \cite{ref36}.  This motivates a systematic investigation of lightweight AI methodologies tailored specifically for UAV-mounted RIS architectures.

Hence, this paper surveys lightweight AI techniques for UAV-mounted RIS systems and provides comparative insights into their performance and complexity characteristics. Also, it presents a case study that highlights the importance of these techniques.
The remainder of this paper is organized as follows. Section \ref{light} explores different lightweight AI techniques used in UAV-mounted RIS networks. Section \ref{apps} lists important application scenarios for lightweight AI in aerial RIS networks. Section \ref{research_challenges} discusses open research challenges and future directions.  Section \ref{case_study} introduces a case study that shows the effect of MAB on the performance of UAV-mounted RIS scenario. Finally, Section \ref{sec:conclusion} concludes the paper.
\begin{table*}[!htbp]
\centering
\caption{Comparison of Lightweight AI Techniques for UAV-Mounted RIS Systems}
\label{tab:lightai}
\resizebox{0.9\textwidth}{!}{
\begin{tabular}{|c|c|c|c|c|c|}
\hline
\textbf{Category} & \textbf{Technique} & \textbf{Computational Complexity} & \textbf{Energy Efficiency} & \textbf{Adaptability} & \textbf{Onboard Feasibility} \\ \hline

Model Efficiency 
& Pruning 
& Low  
& High 
& Medium 
& High \\ \hline

Model Efficiency 
& Quantization 
& Very Low 
& Very High 
& Medium 
& Very High \\ \hline

Model Efficiency 
& Knowledge Distillation 
& Low 
& High 
& Medium–High 
& High \\ \hline

Model Efficiency 
& TinyML 
& Very Low 
& Very High 
& Medium 
& Very High \\ \hline

Adaptive Decision-Making 
& Lightweight RL/DRL 
& Medium 
& Medium 
& High 
& High \\ \hline

Adaptive Decision-Making 
& Meta-Learning 
& Medium 
& Medium 
& Very High 
& Medium \\ \hline

Adaptive Decision-Making 
& Transfer Learning 
& Medium 
& Medium 
& Very High 
& Medium \\ \hline

Adaptive Decision-Making 
& MAB 
& Low 
& High 
& Medium 
& Very High \\ \hline

Adaptive Decision-Making 
& Edge Computing 
& Medium 
& High 
& High 
& High \\ \hline

Topology-Aware Distributed 
& Federated Learning 
& Medium 
& Medium 
& High 
& Medium \\ \hline

Topology-Aware Distributed 
& Sparse GNN 
& Medium 
& Medium 
& High 
& Medium \\ \hline

Hybrid Model-Driven 
& Deep Unfolding 
& Low–Medium 
& High 
& High 
& High \\ \hline

\end{tabular}
}
\end{table*}

\section{Summary of Lightweight AI Techniques for UAV-Mounted RIS Systems}
\label{light}
The principal lightweight AI methodologies applicable to UAV-mounted RIS systems are divided into four main categories as follows:- 

\subsection{Model Efficiency techniques}
\subsubsection{Model Compression: Pruning and Quantization}
Model Compression significantly reduces NN complexity while preserving performance. Pruning eliminates redundant weights or filters, reducing memory access and inference latency. Quantization reduces numerical precision, thereby lowering computational cost and energy consumption.
In UAV-mounted RIS systems, compressed DRL or BeamForming (BF) models can operate efficiently on embedded processors. These techniques are especially important when controlling large-scale RIS panels, where inference complexity scales with the number of reflecting elements. Model compression enhances onboard feasibility without sacrificing adaptability.

\subsubsection{Knowledge Distillation}
It transfers knowledge from a large, computationally intensive "teacher" model to a smaller "student" network suitable for resource-constrained deployment. In UAV-mounted RIS systems, most existing work relies on DRL or complex joint-optimization frameworks \cite{ref7,ref22,ref24}. While these approaches achieve strong performance, their computational overhead may limit real-time onboard implementation.
Large offline-trained DRL models developed at ground stations could be distilled into compact student models for embedded aerial controllers. This would enable near-optimal trajectory and phase configuration control with significantly reduced inference complexity.

\subsubsection{ Tiny Machine Learning (TinyML) and Onboard Edge Intelligence}
TinyML enhances aerial RIS intelligence, enabling lightweight phase prediction, channel classification, and simple trajectory adaptation. It focuses on deploying ML models on ultra-low-power micro-controllers with minimal memory and processing requirements \cite{ref45}. In UAV-mounted RIS systems, most existing AI-driven optimization frameworks rely on DRL or centralized training approaches \cite{ref7,ref22,ref24}, which achieve strong performance in simulation environments but may be practically computationally intensive.  By eliminating reliance on centralized processing, TinyML reduces communication overhead and latency. This is beneficial in emergency deployments or NTN-integrated scenarios where backhaul connectivity may be limited. Given the strict energy and payload constraints of UAV platforms \cite{ref6,ref21}, TinyML offers a practical pathway toward fully autonomous UAV-mounted RIS systems. 

\subsection{Adaptive Decision making methods}
\vspace{-2mm}
\subsubsection{RL and DRL}
RL has become a dominant framework for UAV-mounted RIS optimization due to its ability to handle model-free and dynamic control problems. RL-based methods jointly optimize UAV trajectories and RIS phase configurations via environmental interactions \cite{ref22,ref25,ref28}. DRL extends this capability using neural function approximators to scale to larger state spaces \cite{ref7,ref20,ref21,ref23,ref24}. To enable onboard deployment, lightweight DRL variants reduce network depth, discretize action spaces, and compress state representations. Energy-aware reward formulations further integrate battery constraints into learning policies \cite{ref21,ref34}. 

\subsubsection{ Meta-Learning and Transfer Learning}
Meta-learning enhances generalization by enabling rapid adaptation to new environments with minimal retraining. Given the dynamic propagation conditions of UAV-mounted RIS systems, Meta-RL frameworks are particularly effective \cite{ref14}. Transfer learning techniques reuse pretrained DRL policies in new deployment scenarios, significantly reducing computational overhead and training time \cite{ref33}.
These approaches are valuable in heterogeneous environments such as maritime communications, NTN-integrated systems, and urban mobility scenarios.

\subsubsection{Multi-Armed Bandit (MAB) Methods}
MAB algorithms provide a computationally efficient stateless RL framework. 
It enables fast decision-making with minimal overhead. In UAV-mounted RIS systems, MAB has been applied to mmWave beam selection, RIS element allocation, and WiGig enhancement \cite{ref36,ref39,ref40}. Energy-aware MAB schemes further incorporate battery constraints \cite{ref37,ref38}. MAB techniques are well-suited to real-time onboard implementation.

\subsection{Topology-aware distributed methods}
\subsubsection{Sparse Graph Neural Networks (GNNs)}
GNNs effectively model interactions among UAVs, users, and RIS elements. However, full GNN implementations may be computationally expensive. Sparse GNN architectures reduce message-passing operations by limiting neighborhood connectivity. In multi-UAV RIS deployments \cite{ref32}, sparse GNNs enable scalable topology-aware resource optimization while maintaining manageable computational cost. These approaches are suitable for both cooperative and distributed aerial networks.

\subsubsection{Federated Learning}
FL supports decentralized model training across multiple UAV nodes without sharing raw data \cite{ref31}. This reduces communication overhead and enhances privacy. In cooperative UAV-mounted RIS systems \cite{ref32}, FL enables distributed optimization of trajectories and phases while reducing the signaling overhead. FL is particularly effective in large-scale aerial swarms or NTN-integrated architectures.

\subsubsection{Edge Computing}
Instead of performing full-scale trajectory and phase optimization onboard the UAV, edge-assisted architectures allow real-time coordination between aerial RIS platforms and ground-based MEC units \cite{ref26,ref27}. This reduces onboard processing burden, lowers energy consumption, and improves scalability. Recent work has further investigated EE UAV-mounted RIS-assisted MEC systems, in which joint optimization of communication, computation, and reflection control is performed to maximize overall system efficiency \cite{ref10}. By combining edge computing with lightweight AI techniques, UAV-mounted RIS networks can achieve low-latency decision-making while preserving battery life and maintaining high adaptability in dynamic 6G environments.
\subsection{Hybrid model-driven learning}
\subsubsection{Deep Unfolding}
It integrates iterative optimization algorithms into the layers of structured NNs. Each layer corresponds to an iteration of a classical optimization algorithm, preserving interpretability while reducing parameter count \cite{9524496}. For UAV-mounted RIS phase optimization, deep unfolding provides a balance between model-based precision and data-driven flexibility. Compared with generic DNNs, unfolding architectures have lower computational complexity and faster convergence, making them suitable for embedded UAV controllers.

\vspace{-1.5mm}
\section{Application Domains of Lightweight AI in UAV-Mounted RIS Systems}
\label{apps}
This section discusses the major application domains in which lightweight AI techniques are employed to optimize UAV trajectory, RIS phase configuration, energy management, and distributed coordination. Table~\ref{tab:AI_UAV_RIS} summarizes recent application domains and utilized AI solutions.
\begin{table*}[!t]
\centering
\caption{Summary of AI-Based Techniques for UAV-Mounted RIS Optimization}
\label{tab:AI_UAV_RIS}
\resizebox{0.9\textwidth}{!}{
\begin{tabular}{|c|c|c|c|c|c|}
\hline
\textbf{Ref.} & \textbf{Application Scenario}& \textbf{AI Technique} & \textbf{Optimization Target} & \textbf{RIS Type} & \textbf{Energy-Aware}   \\ \hline

\cite{ref7}  & mmWave-NOMA& DRL & Trajectory + Phase Control & Passive & No   \\ \hline
\cite{ref20} & ISAC Systems&DRL & ISAC Optimization & Hybrid & Yes  \\ \hline
\cite{ref21} & IoT Networks&DRL & Energy-Efficient Phase Control & Passive & Yes   \\ \hline
\cite{ref22} & General UAV-RIS& RL & Trajectory + Phase Shift & Passive & No   \\ \hline
\cite{ref23} & AoI Optimization &DRL & AoI-Aware STAR-RIS Control & STAR-RIS & No  \\ \hline
\cite{ref24} & CoMP-NOMA& DRL & Trajectory + Phase Optimization & Passive & No   \\ \hline
\cite{ref25} & Secure 6G Networks& RL & UAV Swarm Formation & Passive & Yes   \\ \hline
\cite{ref30} & Low-Altitude Economy&DRL & SWIPT + Security Optimization & Active & Yes  \\ \hline
\cite{ref31} & Distributed UAV Networks& Federated Learning & Trajectory + Phase Aggregation & Passive & Partial  \\ \hline
\cite{ref32} & Multi-UAV Systems& Learning-Based Optimization & Cooperative Resource Allocation & Passive & Yes   \\ \hline
\cite{ref33} & ISAC Systems& Transfer Learning + DRL & Hybrid-RIS ISAC Control & Hybrid & No   \\ \hline
\cite{ref34} & Maritime Communications &DRL & Energy Harvesting + Anti-Jamming & Passive & Yes  \\ \hline
\cite{ref35} & Secure Communications & RL & Anti-Jamming Optimization & Passive & No   \\ \hline
\cite{ref36} & mmWave Systems & MAB & Dual-Objective Phase Control & Passive & Yes  \\ \hline
\cite{ref37} & mmWave UAV-RIS& Energy-Aware MAB & Beam Selection & Passive & Yes   \\ \hline
\cite{ref39} & WiGig Networks & Contextual MAB & WiGig Enhancement & Passive & No \\ \hline
\cite{ref40} &  Aerial RIS Systems & Advanced MAB & Phase Selection & Passive & No \\ \hline
\cite{ref14} & Aerial Active-RIS& Meta-RL & Resource Allocation & Active RIS & No   \\ \hline

\end{tabular}
}
\end{table*}

\subsubsection{\textbf{Simultaneous Wireless Information and Power Transfer (SWIPT) and Energy Harvesting (EH)}}
SWIPT, when combined with UAV-mounted RIS, enables joint data transmission and EH. Active and multifunctional RIS architectures are particularly beneficial in such scenarios, as they enhance both the reflected-signal strength and the harvested EE \cite{ref29,ref30}. Learning-based RIS element allocation and SWIPT optimization strategies have been proposed to balance throughput and harvested energy \cite{ref29}. Security-aware SWIPT designs further enhance system resilience while maintaining EE \cite{ref30}. Furthermore, the multi-objective nature of SWIPT systems, which simultaneously optimize data rate, harvested power, and UAV battery sustainability, results in high-dimensional, non-convex optimization problems. Therefore, Lightweight AI techniques are essential to enable real-time control under strict energy constraints while maintaining balanced system performance.

\subsubsection{\textbf{Mobile Edge Computing (MEC)}}
MEC enhances UAV-mounted RIS systems by enabling computational offloading to nearby edge servers. Instead of executing all optimization and learning tasks on board, the UAV computing platform collaborates with MEC servers, thereby offloading computationally intensive tasks to edge infrastructure to reduce latency and processing overhead \cite{ref26,ref27}. More recent frameworks jointly integrate UAV trajectory planning, computation task scheduling, and reflection control, in which the phase configuration of RIS elements is dynamically optimized to manipulate the wireless propagation environment within energy-efficient UAV-mounted RIS-assisted MEC systems. Thus, MEC acts as a system-level enabler that supports lightweight AI and optimization algorithms in computationally demanding scenarios.


\subsubsection{\textbf{Satellite and NTN Integration}}
The integration of UAV-mounted RIS with satellite systems supports NTN architectures and Space–Air–Ground Integrated Networks (SAGIN). In such configurations, UAV-RIS platforms assist LEO satellite communication by improving beam alignment and mitigating propagation loss \cite{ref17,ref18,ref19,ref43}.  Joint optimization across satellite beams, UAV mobility, and RIS phase control yields highly coupled, computationally intensive problems. The presence of Doppler shifts and large-scale MIMO beamforming further increases complexity. Lightweight AI techniques are crucial to ensure real-time feasibility and scalable coordination between satellite and aerial layers while maintaining manageable onboard complexity.

\subsubsection{\textbf{Maritime Communications}}
It is characterized by sparse infrastructure, long propagation distances, and vulnerability to interference and jamming. UAV-mounted RIS systems enhance signal coverage by dynamically adjusting aerial positioning and reflective BF \cite{ref12}. However, the dynamic sea environment introduces rapidly varying channel conditions and Doppler effects. Lightweight RL and energy-aware optimization frameworks have been proposed to jointly optimize UAV trajectory and RIS PSs for secure maritime sensing and communication \cite{ref13}. Anti-jamming and EH-based DRL strategies further enhance robustness and sustainability in maritime deployments \cite{ref34}. Hence, lightweight AI is essential to enable adaptive BF and secure control while preserving UAV battery life during long-duration missions.

\subsubsection{\textbf{ISAC}}
It introduces additional complexity because sensing objectives (e.g., detection accuracy and radar resolution) must be balanced with communication metrics (e.g., throughput and latency). This multi-objective optimization significantly increases computational requirements. Lightweight AI techniques are critical for enabling real-time ISAC control while preserving EE and onboard feasibility. UAV-mounted RIS platforms are particularly suitable for ISAC due to their aerial mobility and programmable electromagnetic response. By intelligently adjusting the UAV's trajectory and the RIS phase configuration, UAV-RIS systems can simultaneously enhance communication throughput and sensing accuracy. Recent studies have investigated DRL-based EE UAV-RIS-assisted ISAC systems, where BF, sensing accuracy, and UAV trajectory are jointly optimized under energy constraints \cite{ref20}. Hybrid-RIS-enabled UAV-assisted ISAC frameworks leveraging transfer-learning-based DRL have also been proposed to improve adaptability across dynamic environments \cite{ref33}. 
\vspace{-1.5mm}
\section{Challenges and Future Directions}
\label{research_challenges}
Despite rapid progress in UAV-mounted RIS systems and AI-driven optimization, several fundamental challenges remain open. In particular, integrating lightweight AI with aerial mobility, programmable reflection\footnote{The phase shifts of RIS elements are dynamically adjusted to control signal propagation} and energy-constrained platforms poses unique challenges that warrant further investigation.

\subsubsection{\textbf{Scalability with Large-Scale RIS Arrays}}
As the number of RIS elements increases, the dimensionality of the optimization problem grows significantly. Jointly with communication metrics (e.g., throughput and latency) and scheduling, it quickly becomes computationally prohibitive. Most existing work relies on DRL, which scales poorly as state and action dimensions increase. Hence, lightweight AI solutions such as pruning, quantization, and model-driven learning have not yet been systematically integrated into large-scale RIS optimization. Developing scalable architectures that maintain performance while controlling computational complexity remains a critical research challenge.

\subsubsection{\textbf{Real-Time Adaptation under High Mobility}}
UAV-mounted RIS systems operate in highly dynamic environments with time-varying cascaded channels, Doppler shifts, and fluctuating interference. Although RL and MAB frameworks can adapt to environmental changes, convergence speed and inference latency remain major concerns for real-time deployment.
Future research should focus on ultra-fast adaptation through meta-learning that can facilitate rapid policy updates when channel conditions or user distributions change, significantly reducing retraining overhead.
Developing lightweight online learning algorithms with bounded computational complexity is equally important to ensure real-time feasibility on resource-constrained aerial RIS platforms. 





\subsubsection{\textbf{Energy-Constrained Intelligence}}
Aerial RIS platforms are severely constrained by energy and payload limits. Although energy-aware RL/MAB methods exist, unified frameworks that jointly optimize propulsion, communication, battery dynamics, and RIS control are still lacking. Future research must develop integrated trajectory–battery–reflection optimization models that explicitly address the energy–computation trade-off. Moreover, combining SWIPT and EH with lightweight AI requires multi-objective learning to balance throughput, harvested energy, and propulsion cost, supported by battery-aware policies, adaptive power allocation, and dynamic energy budgeting for sustainable UAV operation.






\subsubsection{\textbf{Hardware-Aware and Edge-Assisted Learning}}
 Most existing studies assume ideal processing capabilities and abundant computational resources. However, UAV-mounted RIS controllers operate on embedded processors with limited memory, restricted computational throughput, and strict power budgets. Consequently, effective deployment demands hardware-aware AI design that explicitly accounts for processing latency, memory footprint, and energy consumption. Key research challenges include developing TinyML-based phase-control mechanisms suitable for microcontroller-level deployment, designing quantized and compressed DRL policies for efficient embedded inference, and developing adaptive edge–cloud collaboration strategies that intelligently partition computation between onboard processors and MEC infrastructure. 
 Co-designing hardware and learning algorithms is imperative to bridge the gap between simulation-driven research and real-world UAV-mounted RIS implementation, enabling scalable, reliable, and EE aerial intelligent surfaces for future 6G systems.








\subsubsection{\textbf{Multi-UAV and Distributed Coordination}}
Future 6G networks are expected to deploy multiple UAV-mounted RIS platforms operating cooperatively to extend coverage, enhance reliability, and improve spectral efficiency. However, distributed coordination among multiple aerial RIS units introduces substantial challenges, including synchronization, signaling overhead, limited backhaul capacity, and system stability under dynamic mobility conditions. The coupling between trajectory planning, RIS phase configuration, and inter-UAV interference further increases optimization complexity in multi-agent environments. Promising research directions include FL frameworks that operate under intermittent connectivity and limited bandwidth, Sparse GNN for topology-aware, scalable coordination, and distributed MAB algorithms for low-complexity resource allocation. 
Achieving scalable, resilient, and EE coordination among multiple aerial RIS platforms remains a fundamental open problem for practical large-scale deployment \cite{ref38}.








\subsubsection{\textbf{Robustness and Security in Adversarial Environments}}
UAV-mounted RIS systems are inherently vulnerable to security threats such as jamming, spoofing, eavesdropping, and adversarial manipulation, given their open wireless environment and aerial deployment. Although anti-jamming RL strategies have demonstrated promising performance, the robustness and reliability of lightweight AI models under adversarial conditions remain insufficiently explored. In particular, compressed or quantized models may be more susceptible to adversarial perturbations, raising critical concerns for safety-sensitive applications.
Future research must prioritize developing secure, explainable, lightweight AI frameworks that maintain performance under intentional attacks. This includes adversarially robust RL, secure SWIPT and ISAC optimization mechanisms, and resilient distributed learning architectures.

\subsubsection{ \textbf{Standardization and Experimental Validation}}
The gap between theoretical modeling and practical implementation poses a significant barrier to large-scale deployment. Real-world environments introduce non-idealities, including hardware nonlinearities, synchronization errors, channel estimation inaccuracies, and dynamic interference patterns, which are often overlooked in simulation-based evaluations.
Future research must therefore prioritize prototype development of embedded RIS controllers suitable for UAV integration, along with real-time field trials in challenging environments such as maritime and NTN-integrated scenarios. The creation of standardized benchmark datasets tailored to lightweight UAV-RIS learning is also crucial for fair performance evaluation and reproducibility. In parallel, efforts to standardize aerial RIS architectures, control protocols, and interoperability frameworks will be necessary to ensure ecosystem compatibility. Bridging the gap between simulation and real-world experimentation is essential to translating lightweight AI-based UAV-mounted RIS frameworks into reliable, commercially viable 6G solutions.

\subsubsection{\textbf{Integration with Orthogonal Time Frequency Space (OTFS) and High-Mobility Waveforms}}
Recent studies have demonstrated the effectiveness of OTFS modulation in highly dynamic UAV communication scenarios, particularly for mobility-aware handover and high-speed communication systems \cite{refOTFS1,refOTFS2}.
While OTFS-enabled UAV networks have been investigated, the integration of OTFS with UAV-mounted RIS systems remains largely unexplored. In UAV-mounted RIS architectures, cascaded BS–RIS–UAV and BS–RIS–user channels introduce additional delay–Doppler coupling, phase-alignment complexity, and reflection-induced channel sparsity that are absent in conventional UAV systems. The joint optimization of RIS phase configuration, UAV trajectory, and delay–Doppler-domain resource allocation under OTFS signaling, therefore, leads to a new class of tightly coupled, high-dimensional optimization problems. 
Addressing these challenges requires developing lightweight AI frameworks capable of delay–Doppler-aware BF, exploiting sparse channels, and enabling real-time trajectory adaptation under strict onboard energy and computational constraints.

\section{Case Study: MAB schemes for UAV-mounted RIS}
\label{case_study}
Consider a UAV-mounted RIS that is deployed to assist a mmWave BS in serving multiple distributed hotspots. The UAV-mounted RIS dynamically moves between hotspots to enhance signal coverage for users experiencing non-line-of-sight conditions. Each hotspot has uncertain traffic demand and varying channel conditions, while the UAV operates under strict battery constraints. The objective is to design a trajectory that maximizes the cumulative achievable data rate across served hotspots while ensuring that the total propulsion and hovering energy does not exceed the UAV’s battery capacity. The challenge lies in making sequential service decisions without full CSI and under limited onboard computational capability.

To address this, the trajectory design problem is reformulated as MAB framework, where each hotspot represents an arm and the UAV acts as the learner. At each decision step, the UAV selects the next hotspot to visit based on observed rewards (achieved data rates) and contextual information, including hotspot location, historical payoffs, and residual energy. Classical Upper Confidence Bound (UCB) strategies prioritize arms with high uncertainty, while Thompson Sampling (TS) explores probabilistically via posterior sampling. Contextual TS (CTS) further enhances performance by incorporating side information into a Bayesian linear reward model, enabling the UAV to adapt its decisions to spatial and traffic features. Compared to DRL, these MAB schemes are computationally lightweight, require no extensive training datasets, and naturally balance exploration–exploitation under real-time constraints, making them suitable for energy-limited UAV-mounted RIS platforms.

\begin{figure*}[!t]
    \centering
        \subfigure[Attainable rate]{\includegraphics[ height=0.45\columnwidth]{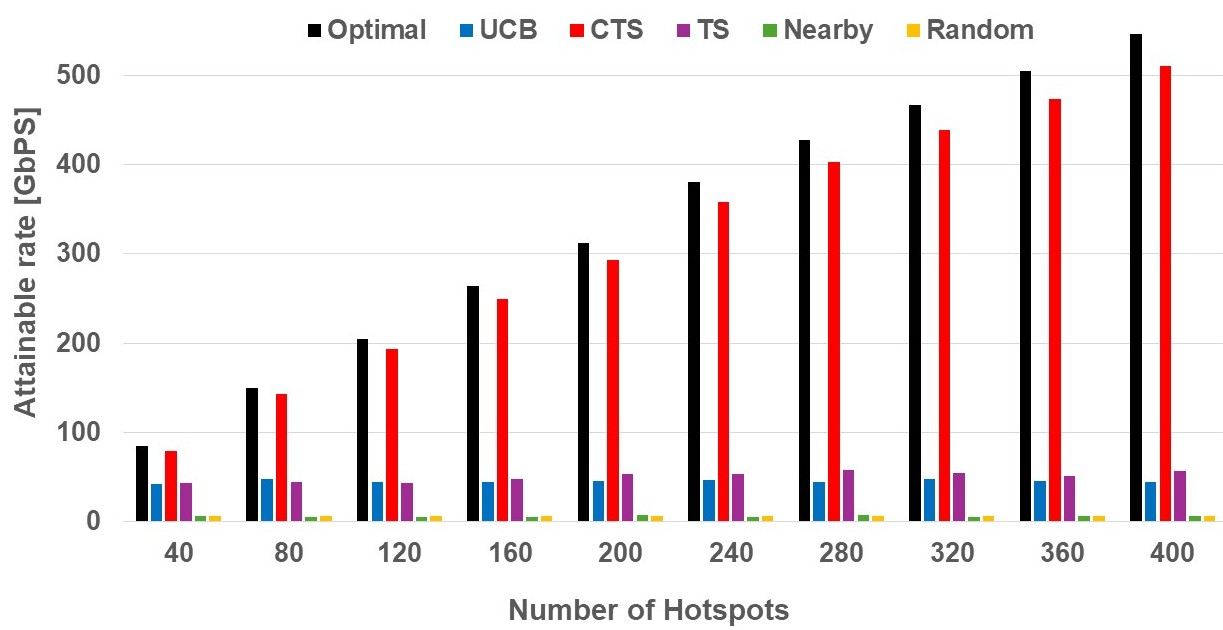}
        \label{throughputvshotspots}
        }
        \qquad
        \subfigure[EE.]{\includegraphics[ height=0.45\columnwidth]{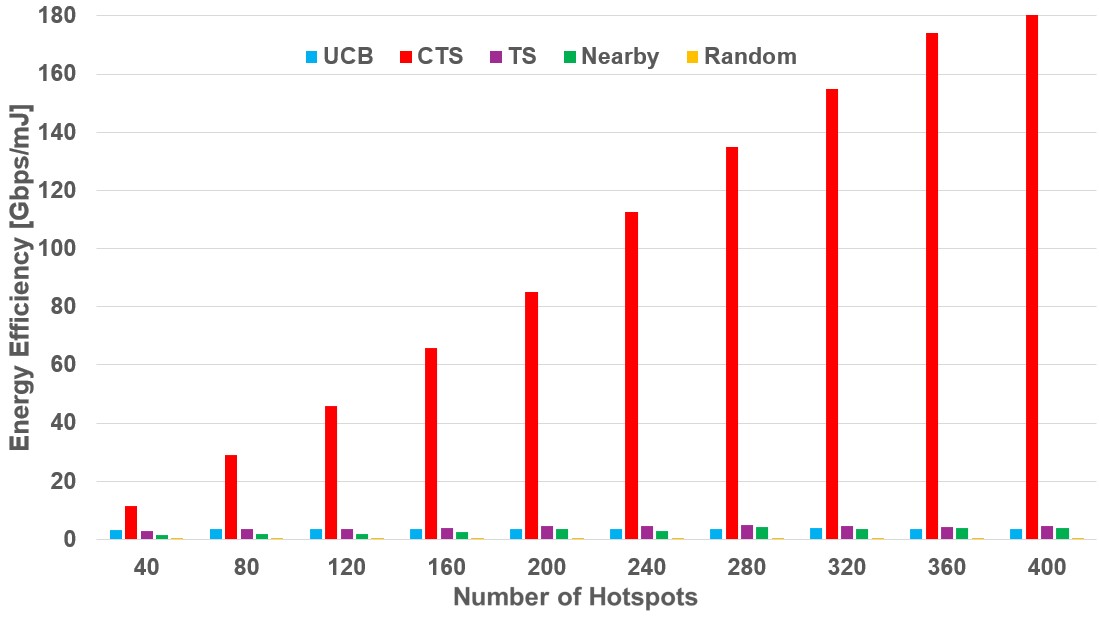}
        \label{EEvshotspots}
        } 
        \vspace{-4mm}
    \caption{Attainable rate and EE Comparison of CTS  against UCB, TS, nearby, and Random schemes at various hotspot numbers and 128 RIS elements within $4\times 4$ $Km^2$ and UAV height of 50m.}
    \label{fig2}
\end{figure*}

Simulation results shown in Fig.\ref{fig2} demonstrate the practical importance of MAB-based lightweight AI. The CTS scheme achieves data rate [Gbps] performance exceeding 90\% of the optimal benchmark while significantly outperforming classical UCB and TS, nearest-hotspot selection, and random policies. In terms of EE [Gbps/mJ], defined as the division of the data rate over
the consumed energy per hotspot, CTS shows orders-of-magnitude improvements over non-contextual or greedy strategies, especially as the number of hotspots increases or the covered region expands. These results confirm that MAB-based decision-making can effectively extend UAV operational lifetime while preserving high communication throughput. Therefore, MAB schemes provide a scalable, low-complexity, and energy-aware intelligence layer for real-time trajectory optimization in airborne RIS systems.

\vspace{-2mm}
\section{Conclusion}
\label{sec:conclusion}
\vspace{-1mm}
This paper provides an overview of lightweight AI techniques for UAV-mounted RIS systems and their role in enabling scalable, EE, and real-time optimization for 6G networks. UAV-mounted RIS enhances coverage, spectral efficiency, and support for emerging applications, including ISAC, SWIPT, NTN integration, and high-mobility communications. However, stringent energy, memory, and computational constraints on UAV platforms necessitate the development of efficient learning frameworks. Hence, we examined lightweight AI techniques, model-efficiency techniques (pruning, quantization, distillation, TinyML), adaptive decision-making methods (RL, meta-learning, MAB), topology-aware distributed learning (FD, sparse GNNs), and hybrid model-driven approaches (deep unfolding) in aerial RIS scenarios. Furthermore, we presented a case study that highlights the importance of the MAB scheme as a lightweight AI for optimizing the trajectory of UAV-mounted RISs with EE considerations. Future work will prioritize hardware–software co-design, embedded implementation, and experimental validation. Integrating UAV-mounted RIS with advanced 6G technologies such as OTFS and distributed edge intelligence will require structured, energy-aware, and delay-sensitive learning architectures. 

\bibliographystyle{IEEEtran}
\bibliography{references}

@ARTICLE{ref2,
  author={Munochiveyi, Munyaradzi and others},
  journal={IEEE Access}, 
  title={{Reconfigurable Intelligent Surface Aided Multi-User Communications: State-of-the-Art Techniques and Open Issues}}, 
  year={2021},
  volume={9},
  number={},
  pages={118584-118605},
  doi={10.1109/ACCESS.2021.3107316}}

@ARTICLE{ref4,
  author={Zhou, Hao and others},
  journal={IEEE Communications Surveys \& Tutorials}, 
  title={{A Survey on Model-Based, Heuristic, and Machine Learning Optimization Approaches in RIS-Aided Wireless Networks}}, 
  year={2024},
  volume={26},
  number={2},
  pages={781-823},
   doi={10.1109/COMST.2023.3340099}}

@ARTICLE{ref5,
  author={Khoshafa, Majid H. and others},
  journal={IEEE Internet of Things Magazine}, 
  title={{Aerial RIS for Enhancing IoT Connectivity: Opportunities and Challenges}}, 
  year={2025},
  volume={},
  number={},
  pages={1-8},
    doi={10.1109/MIOT.2025.3612374}}

@ARTICLE{ref6,
  author={Ahmed, Manzoor and others},
  journal={IEEE Internet of Things Journal}, 
  title={{Toward a Sustainable Low-Altitude Economy: A Survey of Energy-Efficient RIS–UAV Networks}}, 
  year={2025},
  volume={12},
  number={24},
  pages={51951-51975},
   doi={10.1109/JIOT.2025.3618483}}

@ARTICLE{ref7,
  author={Sobhi-Givi, Sima and others},
  journal={IEEE Internet of Things Journal}, 
  title={{Efficient Optimization in RIS-Assisted UAV System Using Deep Reinforcement Learning for mmWave-NOMA 6G Communications}}, 
  year={2025},
  volume={12},
  number={14},
  pages={26042-26057},
   doi={10.1109/JIOT.2025.3553176}}

@ARTICLE{ref8,
  author={Zhao, Jingjing and others},
  journal={IEEE Internet of Things Journal}, 
  title={{Aerial Active STAR-RIS-Aided IoT NOMA Networks}}, 
  year={2025},
  volume={12},
  number={8},
  pages={9525-9538},
   doi={10.1109/JIOT.2025.3531557}}

@ARTICLE{ref9,
  author={Liu, Zhongxu and others},
  journal={IEEE Transactions on Communications}, 
  title={{RIS-Mounted UAV Millimeter-Wave Communications Across Diverse Scenarios: Path-Loss Model, Beam Management, and Posture Analysis}}, 
  year={2026},
  volume={74},
  number={},
  pages={1717-1731},
    doi={10.1109/TCOMM.2025.3637073}}

@ARTICLE{ref10,
  author={Zhang, Haoran and others},
  journal={IEEE Internet of Things Journal}, 
  title={{Energy Efficient Maximization for UAV-Mounted RIS Assisted MEC with Backscatter Systems}}, 
  year={2026},
  volume={},
  number={},
  pages={1-1},
  doi={10.1109/JIOT.2026.3653882}}

@ARTICLE{ref11,
  author={Ye, Yanxin and others},
  journal={IEEE Transactions on Intelligent Transportation Systems}, 
  title={{Trajectory Planning and Transmission Scheduling for UAV-Borne RIS Assisted Energy-Efficient Uplink Transmissions}}, 
  year={2025},
  volume={26},
  number={10},
  pages={17185-17193},
   doi={10.1109/TITS.2025.3602851}}

@ARTICLE{ref12,
  author={Hao, Xu and Ma, others},
  journal={IEEE Transactions on Vehicular Technology}, 
  title={{UAV-Mounted RIS Enabled Maritime Secure Sensing With Joint Beamforming and Trajectory Design}}, 
  year={2025},
  volume={74},
  number={9},
  pages={14833-14837},
   doi={10.1109/TVT.2025.3565951}}

@ARTICLE{ref13,
  author={Lin, Kailong and others},
  journal={IEEE Transactions on Vehicular Technology}, 
  title={{Penalized Reinforcement Learning-Based Energy-Efficient UAV-RIS Assisted Maritime Uplink Communications Against Jamming}}, 
  year={2024},
  volume={73},
  number={10},
  pages={15768-15773},
   doi={10.1109/TVT.2024.3406896}}

@ARTICLE{ref14,
  author={Faramarzi, Sajad and others},
  journal={IEEE Internet of Things Journal}, 
  title={{Meta Reinforcement Learning for Resource Allocation in Aerial Active-RIS-Assisted Networks With Rate-Splitting Multiple Access}}, 
  year={2024},
  volume={11},
  number={15},
  pages={26366-26383},
  doi={10.1109/JIOT.2024.3397007}}

@ARTICLE{ref15,
  author={Goh, Chi Yen and others},
  journal={IEEE Access}, 
  title={{Hybrid Reconfigurable Intelligent Surfaces and Unmanned Aerial Vehicles Assisted Cooperative NOMA}}, 
  year={2026},
  volume={14},
  number={},
  pages={10023-10032},
    doi={10.1109/ACCESS.2026.3652324}}

@ARTICLE{ref17,
  author={Lukito, William D. and others},
  journal={IEEE Internet of Things Journal}, 
  title={{Integrated STAR-RIS and UAV for Satellite IoT Communications: An Energy-Efficient Approach}}, 
  year={2025},
  volume={12},
  number={9},
  pages={11356-11371},
   doi={10.1109/JIOT.2024.3472019}}

@ARTICLE{ref18,
  author={Yao, Wenfei and Chen, Xiaoming and others},
  journal={IEEE Transactions on Communications}, 
  title={{Design of RIS-UAV-Assisted LEO Satellite Constellation Communication}}, 
  year={2025},
  volume={73},
  number={12},
  pages={15656-15671},
   doi={10.1109/TCOMM.2025.3608313}}

@INPROCEEDINGS{ref19,
  author={Yao, Wenfei and others},
  booktitle={2025 IEEE VTC2025-Fall}, 
  title={{Joint Beamforming for RIS-UAV-Assisted LEO Satellite Constellation Communications}}, 
  year={2025},
  volume={},
  number={},
  pages={},
    doi={10.1109/VTC2025-Fall65116.2025.11310163}}

@ARTICLE{ref20,
  author={Chen, Yihui and Yang, Helin and Xie, Wancheng and Lu, Huabing and Zhang, Chenxi},
  journal={IEEE Transactions on Vehicular Technology}, 
  title={{Energy Efficient UAV-RIS-Aided Integrated Sensing and Communication Systems Using Deep Reinforcement Learning}}, 
  year={2026},
  volume={75},
  number={1},
  pages={1613-1618},
   doi={10.1109/TVT.2025.3591011}}

@ARTICLE{ref21,
  author={Salim, Mahmoud M. and others},
  journal={IEEE Communications Standards Magazine}, 
  title={{Energy-Efficient UAV-Mounted RIS for IoT: A Hybrid Energy Harvesting and DRL Approach}}, 
  year={2025},
  volume={},
  number={},
  pages={1-9},
    doi={10.1109/MCOMSTD.2025.3620005}}

@ARTICLE{ref22,
  author={Sun, Tianjiao and others},
  journal={IEEE Transactions on Machine Learning in Communications and Networking}, 
  title={{Reinforcement-Learning-Based Trajectory Design and Phase-Shift Control in UAV-Mounted-RIS Communications}}, 
  year={2025},
  volume={3},
  number={},
  pages={163-175},
    doi={10.1109/TMLCN.2024.3502576}}

@INPROCEEDINGS{ref23,
  author={Zeng, Chenlu and Ke, Feng and Zhang, Xiuyin},
  booktitle={2025 IEEEVTC2025-Spring}, 
  title={{Deep Reinforcement Learning for UAV Assisted AoI-Aware STAR-RIS Communication}}, 
  year={2025},
  volume={},
  number={},
  pages={1-5},
   doi={10.1109/VTC2025-Spring65109.2025.11174562}}

@INPROCEEDINGS{ref24,
  author={Umer, Muhammad and others},
  booktitle={IEEE GLOBECOM 2024}, 
  title={{Deep Reinforcement Learning for Trajectory and Phase Shift Optimization of Aerial RIS in CoMP-NOMA Networks}}, 
  year={2024},
  volume={},
  number={},
  pages={79-84},
   doi={10.1109/GLOBECOM52923.2024.10901709}}

@ARTICLE{ref25,
  author={Tariq, Zain Ul Abideen and others},
  journal={IEEE Access}, 
  title={{RL-Based Adaptive UAV Swarm Formation and Clustering for Secure 6G Wireless Communications in Dynamic Dense Environments}}, 
  year={2024},
  volume={12},
  number={},
  pages={125609-125628},
   doi={10.1109/ACCESS.2024.3455250}}

@ARTICLE{ref26,
  author={Shang, Bodong and others},
  journal={IEEE Wireless Communications}, 
  title={Aerial Reconfigurable Intelligent Surfaces Meet Mobile Edge Computing}, 
  year={2022},
  volume={29},
  number={6},
  pages={104-111},
    doi={10.1109/MWC.001.2200009}}

@ARTICLE{ref27,
  author={Zhai, Zhiyuan and others},
  journal={IEEE Wireless Communications Letters}, 
  title={{Energy-Efficient UAV-Mounted RIS Assisted Mobile Edge Computing}}, 
  year={2022},
  volume={11},
  number={12},
  pages={2507-2511},
   doi={10.1109/LWC.2022.3206587}}

@INPROCEEDINGS{ref28,
  author={Sun, Tianjiao and Yin, Sixing and Li, Jiayue and Li, Shufang},
  booktitle={2024 IEEE ICMLCN}, 
  title={{Optimal Design for Trajectory and Phase-Shift in UAV-Mounted-RIS Communications with Reinforcement Learning}}, 
  year={2024},
  volume={},
  number={},
  pages={101-106},
    doi={10.1109/ICMLCN59089.2024.10624783}}

@INPROCEEDINGS{ref29,
  author={T, Aiswarya and others},
  booktitle={2025 IEEE Conference on Standards for Communications and Networking (CSCN)}, 
  title={{Learning-Based RIS Element Allocation and SWIPT Optimization for Active RIS-Aided UAV Networks}}, 
  year={2025},
  volume={},
  number={},
  pages={},
   doi={10.1109/CSCN67557.2025.11230604}}

@ARTICLE{ref30,
  author={Zhao, Sai and others},
  journal={IEEE Transactions on Vehicular Technology}, 
  title={{Exploit Security for Low-Altitude Economy: A SWIPT-Driven Strategy With UAV-Mounted MF-RIS}}, 
  year={2025},
  volume={},
  number={},
  pages={1-6},
   doi={10.1109/TVT.2025.3599863}}

@INPROCEEDINGS{ref31,
  author={Huang, Chong and others},
  booktitle={IECON 2023}, 
  title={{Federated Learning for RIS-Assisted UAV-Enabled Wireless Networks: Learning-Based Optimization for UAV Trajectory, RIS Phase Shifts and Weighted Aggregation}}, 
  year={2023},
  volume={},
  number={},
  pages={},
  doi={10.1109/IECON51785.2023.10312474}}

@ARTICLE{ref32,
  author={Pan, Hongyang and others},
  journal={IEEE Transactions on Mobile Computing}, 
  title={{Cooperative UAV-Mounted RISs-Assisted Energy-Efficient Communications}}, 
  year={2025},
  volume={24},
  number={10},
  pages={11241-11258},
   doi={10.1109/TMC.2025.3579597}}

@ARTICLE{ref33,
  author={Saikia, Prajwalita and others},
  journal={IEEE Transactions on Communications}, 
  title={{Hybrid-RIS Empowered UAV-Assisted ISAC Systems: Transfer Learning-Based DRL}}, 
  year={2025},
  volume={73},
  number={9},
  pages={8314-8329},
   doi={10.1109/TCOMM.2025.3548797}}

@ARTICLE{ref34,
  author={Yang, Helin and others},
  journal={IEEE Transactions on Wireless Communications}, 
  title={{Energy Harvesting UAV-RIS-Assisted Maritime Communications Based on Deep Reinforcement Learning Against Jamming}}, 
  year={2024},
  volume={23},
  number={8},
  pages={9854-9868},
   doi={10.1109/TWC.2024.3367034}}

@ARTICLE{ref35,
  author={Tariq, Zain Ul Abideen and others},
  journal={IEEE Open Journal of the Communications Society}, 
  title={{Reinforcement Learning for Resilient Aerial-IRS Assisted Wireless Communications Networks in the Presence of Multiple Jammers}}, 
  year={2024},
  volume={5},
  number={},
  pages={15-37},
  doi={10.1109/OJCOMS.2023.3334489}}

@INPROCEEDINGS{ref36,
  author={Hashima, Sherief and others},
  booktitle={2025 IEEE 36th PIMRC}, 
  title={{Dual Objective MAB Scheme for UAV Mounted IRS in mmWave Communications}}, 
  year={2025},
  volume={},
  number={},
  pages={1-7},
   doi={10.1109/PIMRC62392.2025.11275044}}

@ARTICLE{ref37,
  author={Mohamed, Ehab Mahmoud and others},
  journal={IEEE Wireless Communications Letters}, 
  title={{Energy Aware Multiarmed Bandit for Millimeter Wave-Based UAV Mounted RIS Networks}}, 
  year={2022},
  volume={11},
  number={6},
  pages={1293-1297},
   doi={10.1109/LWC.2022.3164939}}

@article{ref38,
  title={{Distribution of multi mmWave UAV mounted RIS using budget constraint multi-player MAB}},
  author={Mohamed, Ehab Mahmoud and Alnakhli, Mohammad and others},
  journal={Electronics},
  volume={12},
  number={1},
  pages={12},
  year={2022},
  doi={10.3390/electronics12010012},
  publisher={MDPI}
}

@INPROCEEDINGS{ref39,
  author={Hashima, Sherief and others},
  booktitle={2023 IEEE 34th PIMRC}, 
  title={{On Enhancing WiGig Communications With A UAV-Mounted RIS System: A Contextual Multi-Armed Bandit Approach}}, 
  year={2023},
  volume={},
  number={},
  pages={1-7},
   doi={10.1109/PIMRC56721.2023.10293924}}

@INPROCEEDINGS{ref40,
  author={Hashima, Sherief and Hatano, Kohei and Mohamed, Ehab Mahmoud},
  booktitle={2023 IEEE 20th CCNC}, 
  title={{Advanced MAB Schemes for WiGig-Aided Aerial Mounted RIS Wireless Networks}}, 
  year={2023},
  volume={},
  number={},
  pages={469-472},
   doi={10.1109/CCNC51644.2023.10060437}}

@ARTICLE{ref43,
  author={Nguyen, Thang V. and others},
  journal={IEEE Transactions on Aerospace and Electronic Systems}, 
  title={{On the Design of RIS–UAV Relay-Assisted Hybrid FSO/RF Satellite–Aerial–Ground Integrated Network}}, 
  year={2023},
  volume={59},
  number={2},
  pages={757-771},
    doi={10.1109/TAES.2022.3189334}}

@ARTICLE{ref44,
  author={Khan, Wali Ullah and others},
  journal={IEEE Wireless Communications}, 
  title={{Integration of Beyond Diagonal RIS and UAVs in 6G NTNs: Enhancing Aerial Connectivity}}, 
  year={2025},
  volume={32},
  number={3},
  pages={56-63},
    doi={10.1109/MWC.001.2400359}}

@ARTICLE{ref44b,
  author={Khan, Wali Ullah and others},
  journal={IEEE Network}, 
  title={{Beyond Diagonal RIS for 6G Non-Terrestrial Networks: Potentials and Challenges}}, 
  year={2025},
  volume={39},
  number={1},
  pages={80-89},
 
  doi={10.1109/MNET.2024.3480332}}

@ARTICLE{ref45,
  author={Liu, Hongfu and others},
  journal={IEEE Transactions on Vehicular Technology}, 
  title={{Tiny Machine Learning (Tiny-ML) for Efficient Channel Estimation and Signal Detection}}, 
  year={2022},
  volume={71},
  number={6},
  pages={6795-6800},
    doi={10.1109/TVT.2022.3163786}}

@article{refOTFS1,
  title={{OTFS-Based Handover Triggering in UAV Networks}},
  author={Mohamed, Ehab Mahmoud and others},
  journal={Drones},
  volume={9},
  number={3},
  pages={185},
  year={2025},
  doi={doi.org/10.3390/drones9030185}
}

@article{refOTFS2,
  title={{UAV selection for high-speed train communication using OTFS modulation}},
  author={Mohamed, Ehab Mahmoud and Hashima, Sherief},
  journal={Scientific Reports},
  volume={15},
  number={1},
  pages={3343},
  year={2025},
  doi={https://doi.org/10.1038/s41598-024-84354-8}
}

@ARTICLE{9524496,
  author={Jagannath, Anu and others},
  journal={IEEE Transactions on Artificial Intelligence}, 
  title={{Redefining Wireless Communication for 6G: Signal Processing Meets Deep Learning With Deep Unfolding}}, 
  year={2021},
  volume={2},
  number={6},
  pages={528-536},
   doi={10.1109/TAI.2021.3108129}}
\end{document}